\documentstyle[twoside,fleqn,espcrc2,epsf]{article}

\newcommand{\Dd}[1]{\mbox{
   \parbox[b]{0cm}{$D$}\raisebox{1.7ex}{$\leftrightarrow$}$_{\!#1}$}}

\newcommand{\sixth}{\mbox{\small $\frac{1}{6}$}}         
\newcommand{\half}{\mbox{\small $\frac{1}{2}$}}          
\newcommand{\quarter}{\mbox{\small $\frac{1}{4}$}}       
\newcommand{\third}{\mbox{\small $\frac{1}{3}$}}         
\newcommand{\msbar}{\mbox{\tiny $\overline{MS}$}}        

\def\lsim{\mathrel{\rlap{\lower4pt\hbox{\hskip1pt$\sim$}}
    \raise1pt\hbox{$<$}}}                
\def\gsim{\mathrel{\rlap{\lower4pt\hbox{\hskip1pt$\sim$}}
    \raise1pt\hbox{$>$}}}                

\title{
       \vspace{-3.65cm}                                     %
       {\normalsize DESY 02-149}      \\[-0.2cm]            
       {\normalsize Edinburgh 2002/11}\\[-0.2cm]            
       {\normalsize LU-ITP 2002/019}\\[-0.2cm]              
       {\normalsize September 2002}  \\                     
       \vspace{1.78cm}                                      
       Calculation of Moments of Structure Functions%
            \thanks{Plenary talk by R. Horsley at Lat02,
                    Boston, U.S.A.}}                        

\author{M. G\"ockeler%
           \address{Institut f\"ur Theoretische Physik, Universit\"at
                    Leipzig, D-04109 Leipzig, Germany}
                         \hspace{-0.25cm} $^,$\hspace{-0.15cm}
           \address{Institut f\"ur Theoretische Physik, Universit\"at
                    Regensburg, D-93040 Regensburg, Germany},
        R. Horsley%
           \address{School of Physics,
                    University of Edinburgh, Edinburgh EH9 3JZ, U.K.},
        D. Pleiter%
           \address{John von Neumann Institute NIC / DESY Zeuthen,
                    D-15738 Zeuthen, Germany},
        P~E.~L. Rakow%
           $^{\rm b}$,
        A. Sch\"afer%
           $^{\rm b}$
        and
        G. Schierholz%
           $^{\rm d,}$%
           \address{Deutsches Elektronen-Synchrotron DESY,
                    D-22603 Hamburg, Germany}}

\begin{document}

\begin{abstract}
The progress on the lattice computation of low moments of both the
unpolarised and polarised nucleon structure functions is reviewed
with particular emphasis on continuum and chiral extrapolations
and comparison between quenched and unquenched fermions.
\end{abstract}

\maketitle

\setcounter{footnote}{0}


\section{INTRODUCTION}
\label{introduction}

Deep Inelastic Scattering (DIS) experiments, such as
$eN \to eX$ or $\nu N \to \mu^- X$
form an important basis for our knowledge of the structure of hadrons.
In these processes the current probe%
\footnote{A more complete set of structure functions available from
DIS processes is given, for example, in \cite{blumlein96a}.}
(either a neutral current, $\gamma$/$Z^0$, or charged current,
$W^+$/$W^-$) with large space-like momentum $-q^2 \equiv Q^2$
breaks-up the nucleon.
The (inclusive) cross section is then determined by
the structure functions $F_1$, $F_2$ when summing
over beam and target polarisations and, in addition, $F_3$ when using
neutrino beams, and $g_1$, $g_2$ when both the beam and target are suitably
polarised. The structure functions are functions of the
Bjorken variable $x$ ($0 \le x \le 1$) and $Q^2$.
(Another class of structure functions -- the transversity
$h_1$ -- can be measured, in principle, from Drell-Yan type processes
or in certain semi-inclusive processes \cite{barone01a}.)
While the original pioneering discoveries were made over
thirty years ago at SLAC, more recently experiments with
polarised beams have been reported and the field remains very active. 
Recent experiments and proposals, \cite{cooper-sarker97a,doyle98a},
include H1 and Zeus at DESY (unpolarised $F_2$ at small $x$,
\cite{h102a} and $F_3$, \cite{zeus02a}),
Hermes at DESY (polarised $g_1$ and $g_2$, \cite{hermes99a}),
E155 at SLAC (polarised $g_1$, $g_2$, \cite{slac99a}),
Jefferson lab (structure functions in the resonance region,
\cite{armstrong01a,niculescu00a}),
COMPASS at CERN (polarised gluon distribution,
$h_1$, $\Lambda$ matrix elements, \cite{compass96a}), CCFR at Fermilab
(unpolarised $F_3$, \cite{ccfr00a}) and RHIC (spin physics, \cite{rhic00a}).
Recent results are given in the DIS conference series, \cite{dis02}.

A direct theoretical calculation of the structure
functions seems not to be possible (but see
\cite{nachtmann02a,caracciolo00a,capitani99a});
however using the Wilson Operator Product
Expansion (OPE) we may relate moments of the structure functions
to matrix elements of certain operators in a twist or Taylor
expansion in $1/Q^2$. Thus if we define
\begin{eqnarray}
   {\cal O}^{\gamma ; \mu_1\cdots\mu_n} &=&
       \bar{q} \gamma^{\mu_1}
                       i\Dd{}^{\mu_2} \cdots i\Dd{}^{\mu_n} q
                                         \nonumber \\
   {\cal O}^{\gamma\gamma_5 ; \sigma\mu_1\ldots\mu_n} &=&
       \bar{q} \gamma^{\sigma} \gamma_5 
                       i\Dd{}^{\mu_1} \cdots i\Dd{}^{\mu_n} q
                                         \nonumber \\
   {\cal O}^{\sigma\gamma_5 ; \sigma\mu_1\ldots\mu_n} &=&
       \bar{q} \sigma^{\sigma\mu_1} \gamma_5 
                       i\Dd{}^{\mu_2} \cdots i\Dd{}^{\mu_n} q
                                         \nonumber
\end{eqnarray}
then we have the Lorentz decompositions%
\footnote{We use
$\langle \vec{p},\vec{s}| \vec{p}^\prime,\vec{s}^\prime \rangle
= (2\pi)^3 2 E_{\vec{p}} \delta({\vec{p}-\vec{p}^\prime})
\delta_{\vec{s},\vec{s}^\prime}$, $s^2 = -m_N^2$.}
\begin{eqnarray}
      \lefteqn{ \langle N(\vec{p}) 
             | {\cal O}^{\gamma ; \{\mu_1 \cdots \mu_n\}}
                    - \mbox{tr} | N(\vec{p}) \rangle =}
      & & 
                                         \nonumber  \\
      & & 2v_n \,
          \left[ p^{\mu_1} \cdots p^{\mu_n} - \mbox{tr} \right]
                                         \nonumber  \\
   \lefteqn{\langle N(\vec{p}, \vec{s}) |
                {\cal O}^{\gamma\gamma_5 ; \{\sigma\mu_1 \cdots\mu_n\}}
                 - \mbox{tr} | N(\vec{p}, \vec{s}) \rangle =}
      & &
                                         \nonumber  \\
      & & 2 {a_n \over n+1} \,
          \left[ s^\sigma p^{\mu_1} \cdots p^{\mu_n} - \mbox{tr} \right]
                                         \nonumber  \\
   \lefteqn{\langle N(\vec{p},\vec{s})|{\cal O}^{\gamma\gamma_5 ;
                             [  \sigma \{ \mu_1 ] \cdots \mu_n \} }
                         - \mbox{\rm tr} | N(\vec{p},\vec{s}) \rangle =}
      & &
                                         \nonumber  \\
      & & {n\over{n+1}} d_n \,
          \left[ (s^{\sigma} p^{\{ \mu_1} - s^{\{ \mu_1} p^{|\sigma|})
                            p^{\mu_2}\cdots p^{\mu_n \}} - \mbox{\rm tr}
          \right]
                                              \nonumber \\
   \lefteqn{\langle N(\vec{p}, \vec{s}) |{\cal O}^{\sigma\gamma_5;
             \sigma\{\mu_1 \cdots\mu_n\}} - \mbox{tr} 
                                       | N(\vec{p}, \vec{s}) \rangle =}
      & &
                                         \nonumber  \\
      & & {2\over m_N} t_n \,
               \left[ (s^\sigma p^{\mu_1}-s^{\mu_1} p^{\sigma})
                                          p^{\mu_2} \cdots p^{\mu_n}
                 - \mbox{tr} \right]
                                         \nonumber
\end{eqnarray}
and the $v_n$, $a_n$, $d_n$ and $t_n$ can be related to moments
of the structure functions. For example we have for $v_n$ and $F_2$
\begin{eqnarray}
   \lefteqn{\int_0^1 dx x^{n-2} F_2(x,Q^2) =}
      & &                                   \nonumber \\
      & & \sum_f E^{(f)\msbar}_{F_2;n}( \mu^2/Q^2, g^{\msbar} )
            v_n^{(f)\msbar}(\mu)
           + O(1/Q^2)
                                         \nonumber
\end{eqnarray}
and similar relations hold between $g_1$ and $a_n$; $g_2$
and a linear combination of $a_n$ and $d_n$; $h_1$ and $t_n$.
Note that $v_n$, $a_n$ (including the $a_n$ part of $g_2$ -- the
so-called Wandzura-Wilczek contribution) and $t_n$ correspond to
twist-2 operators and have a partonic model interpretation%
\footnote{Alternative notations, based on the parton model are
$v_n^{(q)} = \langle x^{n-1} \rangle^q$, $a_n^{(q)} = 2 \Delta^{(n)} q$
and $t_n^{(q)} = 2\delta q^{(n)}$.};
$d_n$ is twist-3 however, and does not have such an interpretation.

Although the OPE gives $v_n$ from $F_1$ (or $F_2$) for even
$n = 2, 4, \ldots$; $v_n$ from $F_3$ for odd $n = 3, 5, \ldots$;
$a_n$ from $g_1$ for $n = 0, 2, \ldots$; $a_n$, $d_n$
from $g_2$ for $n = 2, 4, \ldots$, other matrix elements
can be extracted from semi-inclusive experiments, for
example $a_1$ by measuring $\pi^\pm$ in the final state,
\cite{perutz97a}.

The sum in the previous equation runs over $f = u$, $d$, $s$, $g$, $\ldots$.
We shall only consider $f = u$, $d$ here and mainly the
non-singlet, NS, or proton minus neutron ($p-n$) matrix elements
when the $f = s$ and $g$ (gluon) terms cancel. These latter terms
are less significant for higher moments anyway as the integral
is more weighted to $x \sim 1$ when sea terms have less influence.
The Wilson coefficients, $E^{\msbar}(1,g^{\msbar}(Q))$
are known perturbatively (typically $2-3$ loops).

Present (numerically) investigated matrix elements, 
\cite{gockeler95a,dolgov02a}, are $v_2 \equiv \langle x \rangle$
(which may also be considered as a piece of the momentum sum rule
$\sum_q \langle x \rangle^{(q)} + \langle x \rangle^{(g)} = 1$)
$v_3 \equiv \langle x^2 \rangle$, $v_4 \equiv \langle x^3 \rangle$,
$a_0 \sim \Delta q$ (with a connection to the quark spin component
of the nucleon and also for $\Delta u - \Delta d \equiv g_A$ to
the Bjorken sum rule), $a_1 \sim \Delta q^{(2)}$, \cite{gockeler97a},
$a_2 \sim \Delta q^{(3)}$,
$t_1 \sim \delta q$, \cite{aoki96a,capitani99b},
$t_2 \sim \delta q^{(2)}$, $d_1$ and $d_2$.
We shall mainly discuss here $v_n$ ($n = 1$, $2$, $3$),
$a_0$, $t_0$, $d_1$ and $d_2$. Earlier (lattice conference) reviews
include \cite{martinelli89a,gockeler96a}. Since then emphasis
has been placed first on results with $O(a)$ improved fermions,
considerations of continuum and chiral limits, simulations
with dynamical fermions and recently on the use of chiral
fermions (which can ease the operator mixing problem).
Also possible higher twist contributions and $\pi$, $\rho$ and 
$\Lambda$ matrix elements have been considered.
We shall here briefly review progress in these fields.


\section{THE LATTICE APPROACH}
\label{lattice}

\begin{figure}[htb]
   \vspace*{-0.25in}
   \begin{tabular}{cc}
      \epsfxsize=3.00cm \epsfbox{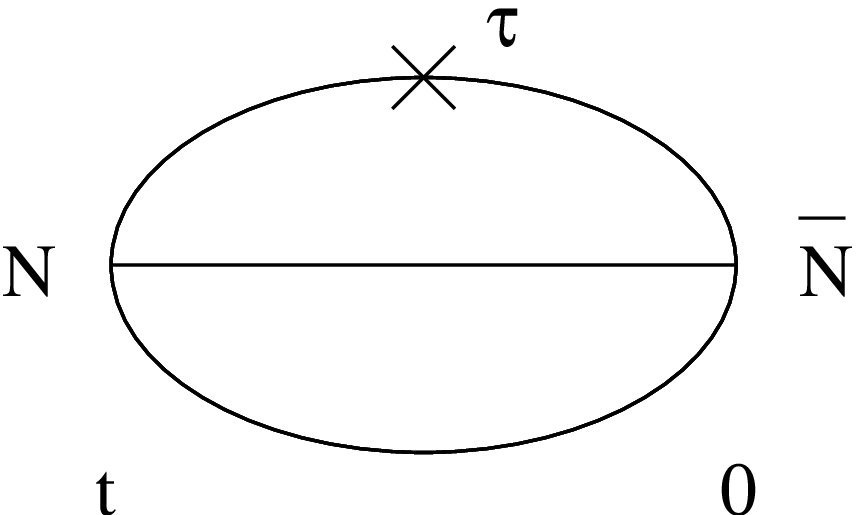}     &
      \hspace{0.50cm}
      \epsfxsize=3.00cm \epsfbox{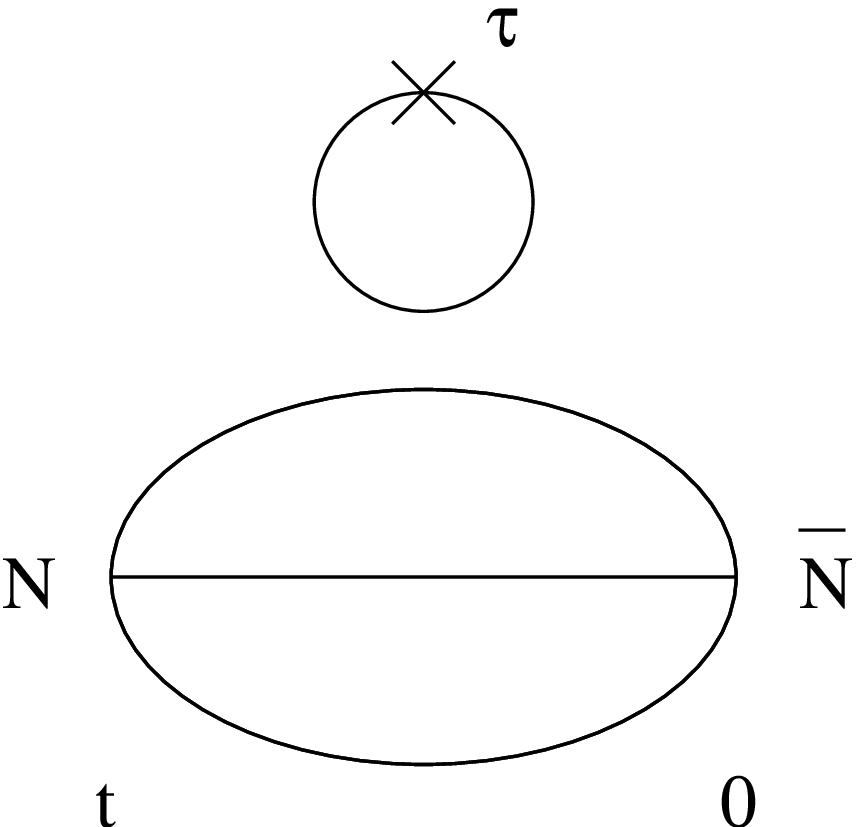}
   \end{tabular}
   \vspace*{-0.30in}
   \caption{\footnotesize{\it The quark-line-connected diagram,
            left hand picture, and quark-line-disconnected diagram,
            right hand picture.
            The cross denotes the operator insertion ${\cal O}(\tau)$.}}
   \vspace*{-0.30in}
   \label{fig_3pt_conn+disconn}
\end{figure}

Matrix elements are evaluated on the lattice, \cite{martinelli89b},
from ratios of (polarised or unpolarised) three-point nucleon correlation
functions to (unpolarised) two-point correlation functions,
\begin{eqnarray}
   R_{\alpha\beta}(t,\tau; \vec{p})
      &=&       {\langle N_\alpha(t;\vec{p}) {\cal O}(\tau)
                   \overline{N}_\beta(0;\vec{p}) \rangle \over
                  \langle N(t;\vec{p}) \overline{N}(0;\vec{p}) \rangle}
                                                \nonumber
\end{eqnarray}
as depicted in Fig.~\ref{fig_3pt_conn+disconn}.
Using transfer matrix methods, it can be shown that
$R \propto \langle N(\vec{p})|{\cal O}|N(\vec{p}) \rangle$ provided that
$0 \ll \tau \ll t \lsim \half N_T$ (the lattice is of size $N_S^3\times N_T$).
As the quark line disconnected diagrams
(RH figure of Fig.~\ref{fig_3pt_conn+disconn}) are difficult
to compute (for some reviews see \cite{okawa95a,gusken99a}),
it is again advantageous to look at non-singlet matrix elements,
such as $v_{n;NS} = v_n^{(u)} - v_n^{(d)} \equiv v_n^p - v_n^n$.
Finally most computations have been carried out in the
quenched approximation, when the fermion determinant
in the partition function is ignored.
This is simply much cheaper in CPU time, but, as will be discussed later,
unquenched results are beginning to appear.

\begin{figure}[htb]
   \vspace*{-0.25in}
   \epsfxsize=7.00cm \epsfbox{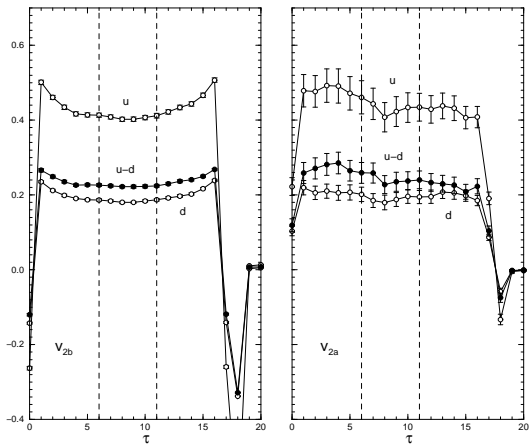}
   \vspace*{-0.30in}
   \caption{\footnotesize{\it The ratio $R$, normalised so that
            when the conditions $0 \ll \tau \ll t \lsim \half N_T$
            are met, the resulting plateau gives the
            bare matrix element. The picture shows the results
            for quenched $O(a)$-improved fermions at $\beta = 6.20$ and
            $\kappa = 0.1344$ on a
            $N_S^3\times N_T \equiv 24^3\times 48$ lattice.
            $t=17$ and a typical fit range for the plateau
            is taken from $\tau = 6$ to $11$, shown by vertical
            dashed lines. The LH picture shows the `diagonal'
            representation, $v_{2b}$ with a stationary nucleon,
            while the RH picture shows the `off-diagonal'
            representation $v_{2a}$, which requires a
            moving nucleon with (lowest possible) momentum
            $\vec{p} = \vec{p}_1 \equiv (2\pi/N_S, 0, 0)$.
            The empty symbols refer to $u$ and $d$ matrix elements,
            while the full symbols give the NS matrix elements.}}
   \label{fig_Rratio_bare_b6p20kp1344_2pics}
   \vspace*{-0.30in}
\end{figure}

Although the Minkowski matrix elements discussed in
section~\ref{introduction} can be written in a Euclidean form in a
straightforward way, the discretisation onto a hypercubic lattice
is not so restrictive as for the continuum
and thus more representations appear, \cite{gockeler96b}.
For example choosing the operators
\begin{eqnarray}
   O^{(q)}_{v_{2a}} &=& \overline{q}
                  \half [\gamma_4 D_1 + \gamma_1 D_4] q
                                         \nonumber  \\
   O^{(q)}_{v_{2b}} &=& \overline{q}
                        [ \gamma_4 D_4 - \third 
                          ( \gamma_1 D_1 + \gamma_2 D_2 + \gamma_3 D_3 )] q
                                         \nonumber
\end{eqnarray}
both lead to a matrix element determining $v_2$. The first 
representation requires a moving nucleon, while for the second
a stationary nucleon is sufficient.
An example for $R$ for these bare matrix elements is shown
in Fig.~\ref{fig_Rratio_bare_b6p20kp1344_2pics}.
Due to the increase in noise in the signal, it is clearly
advantageous to take a stationary nucleon -- but this is unfortunately
only possible for the lowest moments. 

The operators (or raw results) must be renormalised.
If using $O(a)$-improved Wilson type fermions, one also wishes
to improve the operator. At present these additional operators
are known for local (ie no $D$) and one-link (ie one $D$) operators
\cite{capitani00a}, but not for higher-link operators.
Numerically when known these additional operators
do not seem to be significant, \cite{capitani02a}.
Perturbatively $Z$ is known for all
the local operators, and for the one-link operators, \cite{capitani00a}.
For the higher-link operators only the unimproved results are presently
known. For Ginsparg-Wilson (GW) fermions,
the situation is simpler as the improvement
coefficients are simple numbers, \cite{capitani99c}, while
the renormalisation constants are given in \cite{capitani01a}.
Furthermore renormalisation constants for local operators
for Domain Wall (DW) fermions have also been calculated in \cite{aoki98a}.

As most of the perturbative one-loop coefficients are known,
then tadpole improvement (TI) of the renormalisation constants
is possible. There are several variants, the one we shall use
here is given in \cite{capitani01b}. Here the renormalisation
group invariant form of the renormalisation constant
is directly computed. This can then be converted into the
conventional $\overline{MS}$-scheme.

Non-perturbative renormalisation has also been attempted
using both the Schr\"odinger Functional (SF) method, and the RI-MOM-scheme.
The SF method was developed mainly by the ALPHA collaboration, and
presently in quenched QCD most of the improvement coefficients
and renormalisation constants for $O(a)$-improved Wilson fermions
for the local operators are known, \cite{luscher96a,capitani98a},
while for one-link operators ($v_{2a}$) the renormalisation constant
has been determined for both unimproved and $O(a)$-improved fermions,
\cite{guagnelli99a}. An alternative approach, RI-MOM, based on generalising
the perturbative procedure for the determination of the renormalisation
constants has been applied to local, \cite{martinelli94a,gockeler98a},
and higher link operators, \cite{gockeler98a,capitani01b}, for unimproved
Wilson fermions. For $O(a)$ improved Wilson fermions local \cite{lubicz02a}
and one-link operators \cite{gockeler02a} have been investigated.
For DW fermions $Z$ for local operators have been found in \cite{blum01a}.
For unquenched fermions, very little is known at present, \cite{horsley00a}.
However as we later want to compare quenched and unquenched results,
we shall use for consistency the TI Z (except for $Z_{a_0}$,
\cite{gockeler02a}).

Finally the operator mixing renormalisation structure is partially known. 
There is possible additional mixing with operators of the same dimension
for $v_3$ and $v_4$, \cite{gockeler96b}. Also mixing with lower
dimensional operators occurs, in particular for $d_2$, \cite{gockeler00a},
and $d_1$. These are due to additional chiral non-invariant operators
which occur for Wilson fermions (but not for DW or GW fermions).
This point will be discussed further in section~\ref{np_mixing}.
Also the formalism for the SF method 
has been developed for singlet operators mixing with
gluon operators, \cite{palombi02a}.


\section{CHIRAL AND CONTINUUM EXTRAPOLATIONS}
\label{chiral+cont}

\begin{figure}[htb]
   \vspace*{-0.25in}
   \epsfxsize=7.00cm \epsfbox{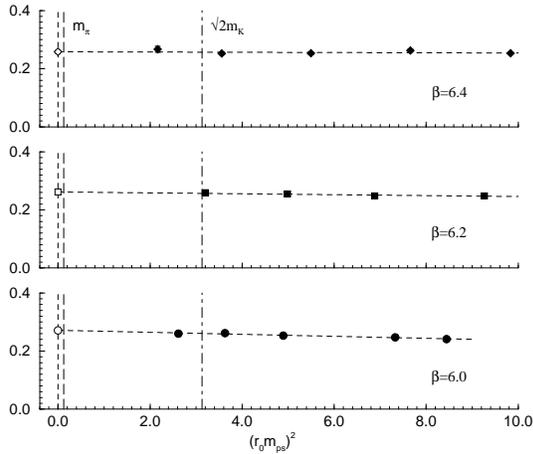}
   \vspace*{-0.30in}
   \caption{\footnotesize{\it $v_{2b;NS}^{\msbar}(2\mbox{GeV})$ versus
                          $(r_0 m_{ps})^2 \sim r_0 m_q$ using
                          $\beta = 6.0$, $6.2$ and $6.4$
                          for $O(a)$-improved quenched Wilson fermions.
                          For orientation, the dash-dotted lines represents
                          (roughly) a strange pseudoscalar meson 
                          ($m_s$ quark mass, determined
                          from $m_K$), and the long-dashed line
                          to the pion ($ud$ quark mass). The chiral
                          limit is given by the short dashed lines.
                          Also shown is a linear extrapolation.}}
   \label{fig_x1b_1u-1d.p0_020613_1251_lat02}
   \vspace*{-0.30in}
\end{figure}

The next step is to attempt a chiral extrapolation. 
An example for quenched $O(a)$-improved Wilson fermions
for $v_{2b}$ is shown in Fig.~\ref{fig_x1b_1u-1d.p0_020613_1251_lat02}.
The points have been scaled to $\overline{MS}$ at $\mu = 2.0\mbox{GeV}$
(using $r_0 = 0.5\mbox{fm} \sim (400\mbox{MeV})^{-1}$).
Also shown is a linear extrapolation to the chiral limit,
\begin{eqnarray}
   v_{n;NS} = a_n (r_0 m_{ps})^2 + b_n
                                              \nonumber
\end{eqnarray}
which seems to be adequate, but one should remember that all the
data points lie at the strange quark mass or higher.
For $v_3$ and $v_4$ similar extrapolations can be performed,
but as they need a moving nucleon yield much more noisy signals.

\begin{figure}[t]
   \vspace*{0.05in}
   \epsfxsize=7.00cm \epsfbox{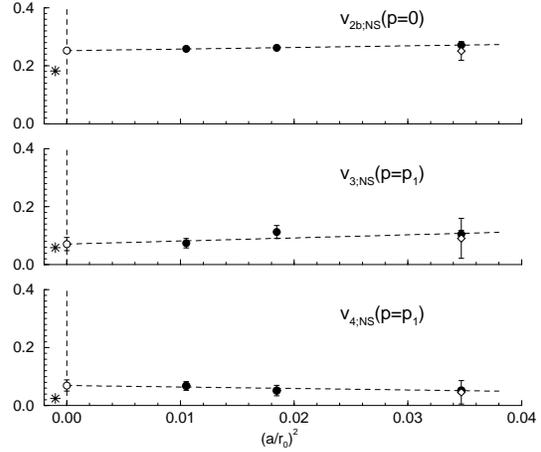}
   \vspace*{-0.30in}
   \caption{\footnotesize{\it The continuum extrapolation
                          for $v_{2b;NS}^{\msbar}$ at a scale of
                          $2\mbox{GeV}$ using the results from
                          Fig.~\ref{fig_x1b_1u-1d.p0_020613_1251_lat02},
                          filled circles with the linear extrapolated
                          result being given by the empty circle.
                          Also shown is the LHPC+SESAM result,
                          \protect\cite{dolgov02a} using unimproved
                          Wilson fermions, empty diamond.
                          The stars are the MRS phenomenological values,
                          \protect\cite{MRS95a}.}}
   \vspace*{-0.30in}
   \label{fig_x1bp0+x2+x3_aor02_MSbar_lat02}
\end{figure}

Finally to obtain the phenomenological result, an extrapolation
to the continuum limit must be performed. These extrapolations
are shown in Fig.~\ref{fig_x1bp0+x2+x3_aor02_MSbar_lat02}
for $v_2$, $v_3$ and $v_4$.
One can see the degradation of the signal as one goes to
the higher moments, which makes the continuum extrapolation
rather noisy. At least for $v_{2b}$ one can say that lattice effects
appear to be small. Also shown is the result from \cite{dolgov02a}
for unimproved Wilson fermions. Good agreement is seen,
which again tends to suggest that lattice effects are small.
The results of the extrapolation are also compared to 
the phenomenological MRS results, \cite{MRS95a}.
It is at present difficult to make any definite statement about the
higher moments $v_3$ and $v_4$ except to say that due
to the continuum extrapolation the ordering has been inverted
(we would expect $v_3 > v_4$). The problem seems to lie in the
continuum extrapolation and can only be cured with more $\beta$-values.
For $v_2$ three $\beta$ values seem to be sufficient;
but the extrapolated result then seems to be about
$30\%$ higher than the phenomenological value.

\begin{figure}[htb]
   \vspace*{-0.20in}
   \epsfxsize=7.50cm \epsfbox{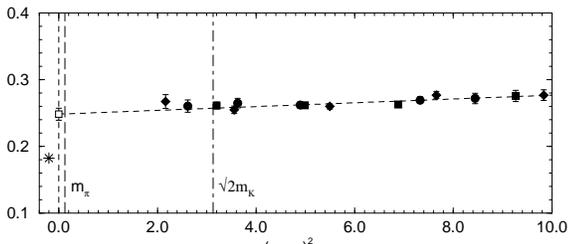}
   \vspace*{-0.50in}
   \caption{\footnotesize{\it $v_{2b;NS}^{\msbar}(2\mbox{GeV})$ with
                          lattice artifacts removed, ie
                          $v_{2b} - c_2 (a/r_0)^2 - d_2ar_0m_{ps}^2$
                          showing the chiral extrapolation for the
                          quenched $O(a)$-improved Wilson results
                          of Figs.~\ref{fig_x1b_1u-1d.p0_020613_1251_lat02}
                          and \ref{fig_x1bp0+x2+x3_aor02_MSbar_lat02}.
                          The same notation as in
                          Fig.~\ref{fig_x1b_1u-1d.p0_020613_1251_lat02}.}}
   \vspace*{-0.30in}
   \label{fig_x1b_1u-1d.p0_020614_1307_lat02}
\end{figure}

One might be worried that one should perform the continuum extrapolation
before the chiral extrapolation. The previous fits can be thought
of as finding the best $\beta$-$(r_0m_{ps})^2$ plane to the data,
so a variant procedure is to try a joint fit, \cite{booth01a},
\begin{eqnarray}
   v_{2b;NS} = a_2 (r_0m_{ps})^2 + b_2 + c_2(a/r_0)^2 + d_2ar_0m_{ps}^2
                                              \nonumber
\end{eqnarray}
where the first two parameters represent the `chiral physics',
the third parameter potential $O(a^2)$ effects and the fourth
parameter $\propto am_q \sim ar_0m_{ps}^2$ is to account for any residual
quark mass effects. (With three $\beta$ values,
one actually reduces the number of free parameters by one.)
In Fig.~\ref{fig_x1b_1u-1d.p0_020614_1307_lat02} we show
the results of this type of fit. The same continuum result is obtained.


\section{UNQUENCHED RESULTS VERSUS QUENCHED RESULTS}
\label{unquenched_results}

\begin{figure}[htb]
   \epsfxsize=7.00cm \epsfbox{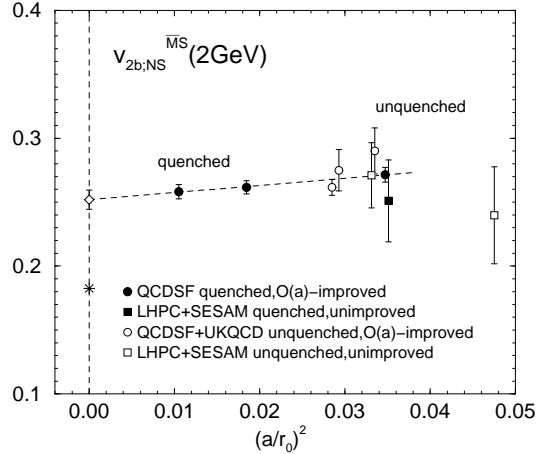}
   \vspace*{-0.30in}
   \caption{\footnotesize{\it Quenched and unquenched results
                          from the QCDSF+UKQCD and LHPC+SESAM Collaborations.
                          Note that to determine $a$ different scales
                          have been used (either $r_0$ or $m_N$ being
                          extrapolated to the chiral limit), and also
                          slightly different renormalisation procedures.}}
   \vspace*{-0.30in}
   \label{fig_x1b_aor02_p0+dyn_MSbar_lat02}
\end{figure}

One possible explanation for the discrepancy between the lattice result
for $v_2$ and the phenomenological result is the use of the
quenched approximation. Indeed one might expect that due to the
momentum sum rule the quenched result is greater than the unquenched result
(as the sea term part is suppressed in the quenched approximation).
While most of the data at present uses quenched fermions,
some recent results using unquenched fermions has appeared:
from the LHPC and SESAM Collaboration, \cite{dolgov02a}
(using unimproved Wilson fermions with $\beta = 5.5$, $5.6$) and
from the QCDSF and UKQCD Collaboration (using $O(a)$-improved
Wilson fermions at $\beta = 5.20$, $5.25$ and $5.29$,
\cite{capitani01b,bakeyev02a}). (Both Collaborations have
three quark mass values at each $\beta$ value.)
Again, as in the quenched case, a linear chiral extrapolation
(at fixed $\beta$) seems adequate.
In Fig.~\ref{fig_x1b_aor02_p0+dyn_MSbar_lat02}
we plot the results against $(a/r_0)^2$.
Are there quenching effects? Although the unquenched results are
not as good as the quenched results, it seems that in this
quark mass ($\gsim m_s$) and $a$ range quenching effects are small.

\begin{figure}[htb]
   \epsfxsize=7.00cm \epsfbox{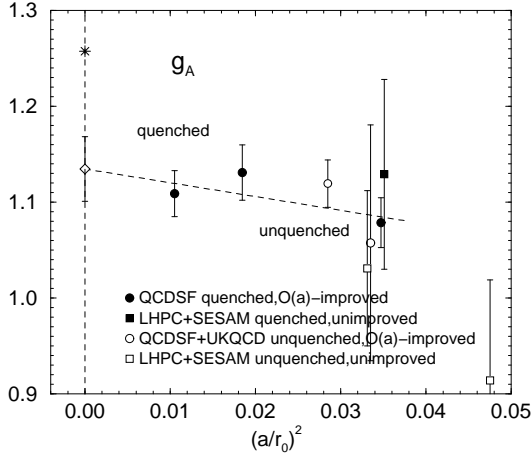}
   \vspace*{-0.30in}
   \caption{\footnotesize{\it Quenched and unquenched results
                          from the QCDSF+UKQCD and LHPC+SESAM Collaborations,
                          for $g_A$ using the same notation as in
                          Fig.~\ref{fig_x1b_aor02_p0+dyn_MSbar_lat02}.}}
   \vspace*{-0.30in}
   \label{fig_ga_aor02_p0+dyn_lat02}
\end{figure}

\begin{figure}[htb]
   \epsfxsize=7.00cm \epsfbox{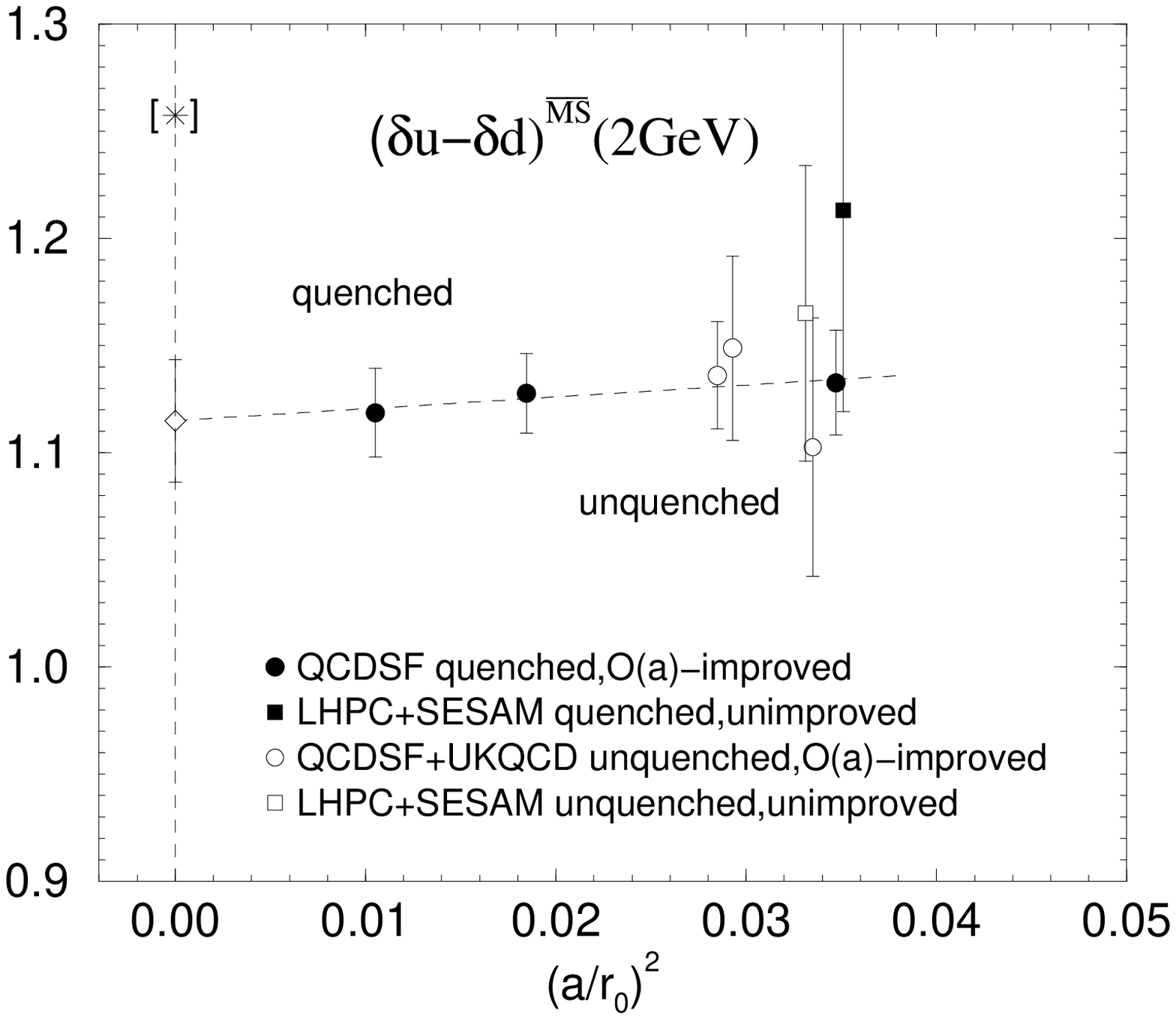}
   \vspace*{-0.30in}
   \caption{\footnotesize{\it Quenched and unquenched results
                          from the QCDSF+UKQCD and LHPC+SESAM Collaborations,
                          for $(\delta u - \delta d)^{\msbar}(\mu)$
                          using the same notation as in
                          Fig.~\ref{fig_x1b_aor02_p0+dyn_MSbar_lat02}.}}
   \vspace*{-0.30in}
   \label{fig_h1b_aor02_p0+dyn+dyn_lat02}
\end{figure}

Further quantities that have been considered include
the axial charge (ie Bjorken Sum rule)
\begin{eqnarray}
   \int_0^1 dx g_1^{p-n}(x,Q^2) 
       &=& \sixth E_{g_1; a_0;NS}\,g_A
                                                \nonumber  \\
   \langle N(\vec{p},\vec{s})| \overline{q} \gamma^\mu \gamma_5 q
                     | N(\vec{p},\vec{s}) \rangle
       &=& 2 s^\mu \Delta q
                                                \nonumber
\end{eqnarray}
with $\Delta u^{\msbar}(\mu) - \Delta d^{\msbar}(\mu) = g_A$
and the tensor charge
\begin{eqnarray}
   \int_0^1 dx h_1^{p-n}(x,Q^2) 
       &=& E^{\msbar}_{h_1; t_0;NS}\,t^{\msbar}_{0;NS}
                                                \nonumber  \\
   \langle N(\vec{p},\vec{s})| \overline{q} i \sigma^{\mu\nu} \gamma_5 q
                     | N(\vec{p},\vec{s}) \rangle
  \hspace*{-0.105in}
       &=& \hspace*{-0.105in} {2 \over m_N} ( s^\mu p^\nu - s^\nu p^\mu )
                      \delta q
                                                \nonumber
\end{eqnarray}
with $\delta u^{\msbar}(\mu) - \delta d^{\msbar}(\mu) = 
t_{0;NS}^{\msbar}(\mu)$.
In Figs.~\ref{fig_ga_aor02_p0+dyn_lat02} and
\ref{fig_h1b_aor02_p0+dyn+dyn_lat02}
we show equivalent pictures to Fig.~\ref{fig_x1b_aor02_p0+dyn_MSbar_lat02}.
Again little difference between the quenched and unquenched simulations
is seen.


\section{TOWARDS SMALL QUARK MASSES}
\label{samm_quark_masses}

The results shown previously have all been characterised by having
data points at quark masses at or above the strange quark mass,
and then a linear extrapolation in the quark mass to the chiral
limit. There has been much recent work developing chiral perturbation
theory, $\chi$-PT, \cite{detmold01a,arndt01a,chen01a,thomas02a}
which has shown the existence of a chiral logarithm of the form
$\sim m_q \ln m_q$,
\begin{eqnarray}
   t_{n;NS} &=& T_n ( 1 - \half C_4 m_{ps}^2 \ln ( m_{ps}^2 / \mu_\chi^2 ) )
                                              \nonumber  \\
   v_{n;NS} &=& V_n ( 1 - C_3 m_{ps}^2 \ln ( m_{ps}^2 / \mu_\chi^2 ) )
                                              \nonumber  \\
   a_{n;NS} &=& A_n ( 1 - C_2 m_{ps}^2 \ln ( m_{ps}^2 / \mu_\chi^2 ) )
                                              \nonumber
\end{eqnarray}
where $C_n = (ng_A^2 + 1) / (4\pi f_\pi)^2$ for (full) QCD. For
quenched QCD an expression for $C_3$ in terms of $F$ and $D$ constants
can be found in \cite{chen01b}. (Note that for quenched QCD
for the nucleon there is no `hairpin' contribution giving rise
to the logarithm $\sim \ln m_q$ for $v_n$.)
What is a suitable value for the chiral scale $\mu_\chi$?
Roughly for $m_{ps} > \mu_\chi$, pion loops are suppressed and
there is a linear variation in $m_q$ ie constituent quark behaviour,
while for $m_{ps} < \mu_\chi$ we have non-linear behaviour.
Often a value for $\mu_\chi \sim 1\mbox{GeV}$ is taken.
We shall also use a comparison value of $500\mbox{MeV}$ here.
From the above formulae, we see that $\chi$-PT
always decreases the value of the matrix element as $m_{ps}^2 \to 0$.
Thus the lattice result should always be larger than the $\chi$-limit
result. Also we would expect more effect for $v_n$ than for $a_n$.
(Recent work in \cite{detmold02a} indicates however that when including
$\Delta$ as well as the $N$ then effectively bending only occurs
for $v_n$ but not $a_n$ and $t_n$.)

\begin{figure}[htb]
   \vspace*{-0.25in}
   \epsfxsize=7.50cm 
   \epsfbox{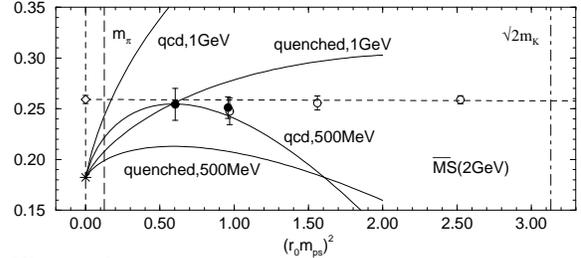}
   \vspace*{-0.45in}
   \caption{\footnotesize{\it Quenched results for unimproved
                          Wilson fermions at $\beta = 6.0$ for
                          $v_{2b;NS}^{\msbar}(2\mbox{GeV})$.
                          The empty circles are for heavier
                          quark masses on a $24^3\times 32$ lattice,
                          $O(200)$ configurations, while the filled
                          circles for the lighter quark masses
                          are on a $32^3\times 48$ lattice,
                          $O(100/50)$ configurations,
                          \protect\cite{gockeler02b}. Finite volume
                          effects are checked at the second lightest
                          mass. The dashed line is
                          a linear fit to the data. From the 
                          MRS value (star), \protect\cite{MRS95a},
                          the $\chi$-PT formula is applied, with
                          $C_3/r_0^2 \sim 0.28$, $0.67$ for quenched or
                          full QCD respectively and for two
                          values of the chiral scale $\mu_\chi$.}}
   \vspace*{-0.30in}
   \label{fig_x1b_1u-1d.p0_020617_1511_lat02}
\end{figure}

As it is not so clear to which quark mass $\chi$-PT is valid, and
as the results shown so far at $m_q \gsim m_s$
yield a linear behaviour it is necessary to
go to lower quark masses. In \cite{capitani01b} this was started
using unimproved quenched Wilson fermions at $\beta = 6.0$
(as the problem of `exceptional configurations' then seems to be less severe).
The present status, \cite{gockeler02b},
is shown in Fig.~\ref{fig_x1b_1u-1d.p0_020617_1511_lat02}.
All quark masses have $am_{ps}N_S \gsim 4$.
As $(r_02m_{\pi})^2 \sim 0.5$, the lightest mass used in the
simulation is $\sim 2m_{ud}$ (this corresponds to $m_{ps}/m_V \sim 0.4$).
Little curvature in the numerical results is seen,
but is still possible, as we expect the coefficient $C_3$
to be smaller than in the unquenched case, \cite{chen01b}.
Also, as noted before, quenching might give a higher value for
$v_{2b}$ than the phenomenological value anyway.

\begin{figure}[t]
   \vspace*{0.10in}
   \epsfxsize=7.50cm 
   \epsfbox{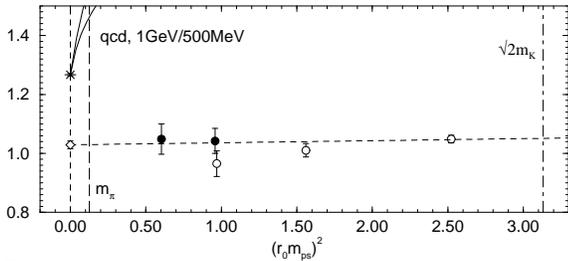}
   \vspace*{-0.45in}
   \caption{\footnotesize{\it Quenched results for unimproved
                          Wilson fermions at $\beta = 6.0$ for
                          $g_A$. Notation as in
                          Fig.~\ref{fig_x1b_1u-1d.p0_020617_1511_lat02}.}}
   \vspace*{-0.30in}
   \label{fig_axial_2_1u-1d.p0_020619_1452_lat02}
\end{figure}

In Fig.~\ref{fig_axial_2_1u-1d.p0_020619_1452_lat02}
we show results for $g_A$.
Note that $\chi$-PT goes in the wrong direction.
It is less clear if there is a finite volume effect.
Ref.~\cite{jaffe01a} suggested that the charge is delocalised
in the chiral (and infinite volume) limit, $g_A \to 2/3 g_A$.
This was further discussed in \cite{cohen01a}, which showed
that for finite volumes, there are no (large) volume effects.

Nevertheless the questions of finite volume effects and the range
of applicability for Wilson fermions near the chiral limit
remain, and recently there have also been results using DW fermions
by the RBC Collaboration. These have much better chiral properties
than Wilson fermions and so are more suitable for investigating
small quark masses. In \cite{orginos02a}, $v_{2b}$ is computed on a
$16^3\times 32 [\times 16]$ lattice at $a^{-1} \sim 1.3\mbox{GeV}$.
Thus the lowest pion mass there, $(am_{ps})^2 \sim 0.1$, corresponds
to about $(r_0m_{ps})^2 \sim 1$ in
Fig.~\ref{fig_x1b_1u-1d.p0_020617_1511_lat02}.
At this point some curvature in the signal is present.
For $g_A$, \cite{ohta02a}, finite volume effects are seen.
Thus the general situation is not completely clear.


\section{OTHER TOPICS}
\label{misc}


\subsection{Non-perturbative mixing}
\label{np_mixing}

A further source of discrepancy between lattice results and phenomenological
results can lie in the incorrect treatment of non-perturbative mixing
of the lattice operators. An example is given by $d_2$, which can 
be found from $g_2$,
\begin{eqnarray}
   \int_0^1 dx x^2 g_2(x,Q^2) 
      \hspace*{-0.1in}&=&\hspace*{-0.1in}
                        \sixth \sum_{q=u,d} \left[ E^{(q)}_{d_2} d_2^{(q)} - 
                                 E^{(q)}_{a_2} a_2^{(q)} \right]
                                            \nonumber  \\
   \int_0^1 dx x^2 g_1(x,Q^2) 
      \hspace*{-0.1in}&=&\hspace*{-0.1in}
                         \quarter\sum_{q=u,d} E^{(q)}_{a_2} a_2^{(q)}
                                            \nonumber
\end{eqnarray}
where $a_2$ and $d_2$ are given in section~\ref{introduction}
as nucleon matrix elements of certain operators
${\cal O}^{(q)}_{a_2/d_2} \sim \overline{q}\gamma\gamma_5 D D q$.
The $a_n$ operators have twist two, but $d_n$ corresponds to twist
three and is thus of particular interest.
A `straightforward' lattice computation,
\cite{gockeler95a,dolgov02a}, gave rather large values
for $d_2^{p}$. A recent experiment, \cite{slac99a},
however indicated that this term was very small. This problem was
traced in \cite{gockeler00a} to a mixing of the original operator
${\cal O}^{(q)}_{d_2}$ with a lower-dimensional operator
${\cal O}_{\sigma}^{(q)} \sim \overline{q} \sigma D q$.
This additional operator mixes $\propto 1/a$ and so its
renormalisation constant must be determined non-perturbatively.
In \cite{gockeler00a} this was attempted using RI-MOM,
and led to results qualitatively consistent with the experimental values.
Note that this is only a problem when using Wilson-like fermions, as
we would expect the operator to appear like
$\sim m_q\overline{q} \sigma D q$ and hence vanish in the chiral limit.
Thus there should be no mixing if one uses GW or DW fermions.
In \cite{orginos02a} this was investigated for $d_1$
using DW fermions and compared with unimproved Wilson fermions $d_1$
results from \cite{dolgov02a}. The same phenomenon was seen: $d_1$ using
DW fermions gave a small value in the chiral limit, while
the unimproved Wilson fermion results increased
strongly as the quark mass was reduced.


\subsection{Higher Twist effects}
\label{higher_twist}

Potential higher twist effects are present in the moment of a structure
function, see section~\ref{introduction}. These $O(1/Q^2)$ terms
have four quark matrix elements. A general problem is the non-perturbative
mixing of these new dimension 6 operators with the previous dimension
4 operators. At present results are restricted to finding combinations
of these higer twist operators which do not mix from flavour symmetry.
In \cite{capitani99d} the lowest moment of the pion structure
function was considered,
\begin{eqnarray}
   \lefteqn{\int_0^1 dx F_2(x,Q^2)|^{I=2}_{Nachtmann} =}
       & &
                                         \nonumber  \\
       & &
        1.67(64) { f_\pi^2 \alpha_s(Q^2) \over Q^2 } + O(\alpha_s^2)
                                            \nonumber
\end{eqnarray}
where the $SU_F(2)$ flavour symmetry group gives the combination
$F_2^{I=2}=F_2^{\pi^+} + F_2^{\pi^-} -2 F_2^{\pi^0}$. For the
nucleon the $SU_F(3)$ flavour symmetry group must be considered,
ie taking mass degenerate $u$, $d$ and $s$ quarks, \cite{gockeler01a},
giving
\begin{eqnarray}
   \lefteqn{\int_0^1 dx F_2(x,Q^2)|^{27,I=1}_{Nachtmann} =} 
       & &
                                         \nonumber  \\
       & &
        -0.0005(5) { m_p^2 \alpha_s(Q^2) \over Q^2 } + O(\alpha_s^2)
                                            \nonumber
\end{eqnarray}
(To access this moment experimentally needs
the measurement of structure functions of $p$, $n$, $\Lambda$, $\Sigma$
and $\Xi$ baryons.)
These results are for quenched unimproved
Wilson fermions at $\beta = 6.0$, and are very small
in comparison with the leading twist result.
However these are rather exotic combinations
of matrix elements and say little about individual contributions. 
Nevertheless this might hint that higher twist contributions are small.


\subsection{Pion, Rho and Lambda results}
\label{pion_rho_lambda}

Moments for pion and rho structure functions were 
computed in \cite{best97a}, for unimproved Wilson fermions.
Using the SF method, $v_{2a}$ was calculated for the pion,
\cite{jansen99a} for both unimproved and $O(a)$-improved fermions,
giving numbers in agreement with \cite{best97a}.
Finally there have been results for moments of $\Lambda$
structure functions, \cite{gockeler02c}. These are potentially useful
as one can compare with nucleon spin structure
and check violation of $SU_F(3)$ symmetry. First indications
are that there is little flavour symmetry breaking.


\section{CONCLUSIONS AND FUTURE PERSPECTIVES}
\label{conclusions}

Clearly the computation of many matrix elements giving low moments of
structure functions is possible. We would like to emphasise that
a successful computation is a fundamental test of QCD -- this
is not a model computation. There are however many problems to overcome:
finite volume effects, renormalisation and mixing, continuum and
chiral extrapolations and unquenching.
At present although overall impressions are encouraging,
still it is difficult to re-produce experimental/phenomenological 
results of (relatively) simple matrix elements (eg $v_2$, $g_A$).
Improvements are thus necessary in all areas. Nevertheless progress
is being made: there are now considerations of both chiral
and continuum extrapolations, some dynamical results are now available,
there are attempts to understand lower quark mass both numerically
(using both Wilson fermions and DW fermions) and from $\chi$-PT.
Clearly everything depends on the data and the quest for better
results should continue. To leave the region where constituent
quark masses give a reasonable description of the data, 
seems unfortunately to require quark masses rather close
to the $ud$ mass. This will presumably also entail the use of
unquenched chiral fermions (such as GW/DW).
This will need much faster machines and is, perhaps, a cautionary 
tale for the determination of other matrix elements.


\section*{ACKNOWLEDGEMENTS}

The QCDSF collaboration numerical calculations were performed
on the Hitachi {\it SR8000} at LRZ (Munich), the APE100,
APEmille at NIC (Zeuthen) and on the Cray {\it T3E}s at
NIC (J\"ulich) and ZIB (Berlin) while the UKQCD collaboration
unquenched configurations were obtained from the Cray {\it T3E}
at EPCC (Edinburgh). This work is supported by the DFG, by BMBF
and by the European Community's Human potential programme
under HPRN-CT-2000-00145 Hadrons/LatticeQCD.



\end{document}